\documentclass[onecolumn,preprintnumbers,amsmath,amssymb,pra]{revtex4}
\usepackage{graphicx}
\usepackage{epsfig}
\usepackage{epsf}
\usepackage{amssymb}
\usepackage{dcolumn}
\usepackage{bm}
\usepackage{subfigure}
\usepackage{setspace}
\usepackage{multirow} 
\usepackage{ulem} 
\usepackage{color} 
\graphicspath{{../../figures/Encoded/}}
\def\ThisFile{\jobname}

\begin{document}
\title{Subsystem Pseudo-pure States}
\author{P. Cappellaro, J.S. Hodges, T.F. Havel and D.G. Cory}
\affiliation{Massachusetts Institute of Technology \\ Department of Nuclear Science and 
Engineering, Cambridge, MA 02139, USA}
\date{\today}
\newcommand{\half}{\frac{1}{2}}
\newcommand{\ket}[1]{\vert{#1}\rangle}
\newcommand{\bra}[1]{\langle{#1}\vert}
\newcommand{\ham}{\mathcal{H}}
\newcommand{\Tr}[1]{Tr\{{#1}\}}
\newcommand{\jh}[1]{\textcolor{red}{\textsf{#1}}}
\newcommand{\pc}[1]{\textcolor[gray]{.3}{\textbf{#1}}}
\newcommand{\jhC}[1]{\textcolor{red}{\textsf{#1}}} 
\newcommand{\pcC}[1]{\textcolor[gray]{.5}{\textsf{#1}}} 

\begin{abstract}
A critical step in experimental quantum information processing (QIP) is to implement control of quantum systems protected against decoherence via informational encodings, such as quantum error correcting codes, noiseless subsystems  and decoherence free subspaces. These encodings lead to the promise of fault tolerant QIP, but they come at the expense of resource overheads.

 Part of the challenge in studying control over  multiple logical qubits, is that QIP test-beds have not had sufficient resources to analyze encodings beyond the simplest ones. The most relevant resources are the number of available qubits and the cost to initialize and control them. Here we demonstrate an encoding of logical information that permits the control over multiple logical qubits without full initialization, an issue that is particularly challenging in liquid state NMR. The method of subsystem pseudo-pure state will allow the study of  decoherence control schemes on up to 6 logical qubits using liquid state NMR implementations.\\
\end{abstract}
\maketitle

\section{Introduction}
Liquid state Nuclear Magnetic Resonance (NMR) is a convenient test-bed for new ideas in Quantum Information Processing (QIP). NMR has been used to experimentally test quantum algorithms \cite{7NMR,ChuangShor15}, control strategies \cite{khanejaOCT,PBEFFHMC:03,ChuangControl,CappellaroHodgesLeakage} and steps toward fault tolerant quantum computation \cite{coryQEC,DFSEvan,LidarExp}. Recently, the focus has turned to control of encoded information that will allow quantum processors to avoid decoherence. The advantages achieved by encoding information come at the expenses of physical resources, as the encoding requires additional qubits as ancillas \cite{shor95,DFSZanardi,DFSGuo}. For liquid state NMR to continue its role as a QIP test-bed, the size of the systems used must increase. This is a critical issue because of the signal loss for pseudo-pure states accompanying each added qubit \cite{PNASalgocooling,ChuangPP}.  Here we propose a new scheme to reduce this signal decrease, that will allow the study of 3-6 logical qubits with the current experimental instrumentation. We first review the sources of signal loss tied to the creation of pseudo-pure states in NMR and introduce a new class of pure states that we call \textit{subsystem} pseudo-pure states,  for which this signal loss is significantly reduced. Before quantifying the gain in signal for a general encoding, we present an example drawn from a widely studied encoding. Finally, we discuss the experimental results with particular attention to the metrics of control that the new kind of subsystem pseudo-pure states allows to measure and their ability to quantify the actual control reached in experimental tests.

\section{Subsystem pseudo-Pure States}
At room temperature and in a high magnetic field, the NMR spin system is a highly mixed state described by the thermal density matrix $\rho_{th}$: 
\begin{equation}
\rho_{th} \approxeq \frac{\openone}{2^N}-\epsilon\rho_{eq}=\frac{\openone}{2^N}-\frac{\epsilon}{2^N}\sum_{i=1}^N\sigma_z^i
\end{equation}
where the term $\epsilon\rho_{eq}$ is a small, traceless deviation from the identity, which gives rise to the observable signal.
The ability to use this system as a quantum information test-bed relies on effectively purifying the mixed equilibrium state. 
 QIP can be performed on \textit{pseudo-pure} states \cite{ChuangPP,Knill,CoryPP}, states for which the dynamics of the observable operators  are equivalent to the observables of a pure state.
Unfortunately, the creation of pseudo-pure states comes at the expense of exponential consumption in experimental resources: time in the case of temporal averaging \cite{Knill}, signal in the case of spatial averaging \cite{CoryPP}, or usable Hilbert space in the case logical labeling \cite{ChuangPP}. 

Since the eigenvalues of a pseudo-pure state are different than those of the mixed state (with the exception of $SU(2)$), a non-unitary completely positive operation, T, must be implemented:  
\begin{equation}\label{pp}\begin{array}{l}
	\rho_P=T(\rho_{th})
	=\frac{\openone}{2^N}-\epsilon\alpha(\rho_{pp}-\frac{\openone}{2^N})
\end{array}\end{equation}
where $\rho_{pp}$ is a density matrix describing a pure state.
The scaling factor $\alpha$ determines the signal loss and is bounded by the spectral norm ratio (since $\left\| \rho_{eq} \right\| \geq \left\|T(\rho_{eq})\right\|$):
\begin{equation}\label{alpha}
	\alpha \leq \frac{\left\| \rho_{eq} \right\|}{\left\|\rho_{pp}-\openone/2^N\right\|}
\end{equation}
with $\left\| \rho_{eq} \right\|=\frac{N}{2^N}$. 
The SNR loss in the case of a full pseudo-pure state is thus $\frac{N}{2^N-1}$. 
This exponential loss of signal is one of the facts that disqualify liquid state NMR as a scalable approach to QIP \cite{CavesNMR, Warrenscience}. Even if algorithmic purification schemes \cite{PNASalgocooling,algocooling,baughcooling} could be applied, if there was a sufficient number of spins over which we had high fidelity control, stronger spin-spin couplings and longer $T_2$ times would be required. Considering we have the ability to coherently control 10-12 qubits \cite{12qubits} and possess spin systems responsive to such control, the loss of signal when preparing pseudo-pure states is also a serious limitation for benchmarking these systems. 

To avoid this SNR loss when studying encoded operations, we can use the additional flexibility afforded since only the subsystems encoding  the information need to be pseudo-pure, while all other subsystems can be left in a mixed state. This also reduces the complexity of the state preparation. In a parallel way \cite{LidarPPS}, it had been observed that imperfect state preparation is enough for the existence of DFS. 
To present the general structure that encoding imposes to the Hilbert space, we adopt the subsystem approach \cite{QECNS,DFSkempe}, that provides a unified description for Quantum Error Correction (QEC) \cite{QEC,CalderbankShor96,steane96}, Decoherence Free Subspaces (DFS) \cite{DFSZanardi,DFSGuo,DFSLidar}  and Noiseless Subsystems (NS) \cite{NSviola,DFSDefilippo,DFSuniversal1,DFSuniversal2}.  A Hilbert space $\mathcal{H}$ of dimension $d=2^N$ is used to encode $l\leq N$ qubits of information, protected against some noise $J=\{ J_\alpha\}$. 
With a change of basis to a direct sum
\footnote{Further action of an appropriate operator $A$ is required in the case of QEC codes \cite{NSviola}} 
$	\mathcal{H} \cong \bigoplus_i \mathcal{L}_i \otimes \mathcal{S}_i  
$, 
 the noise acts only on the subsystems $\mathcal{S}_i$ (the syndrome) while the subsystems $\mathcal{L}_i$ are noiseless (for simplicity,  we will often refer to a decomposition: $\ham=\mathcal{L}\otimes\mathcal{S}\oplus\mathcal{R}$, with $\mathcal{R}$ an unprotected subspace).

To perform computations on logical qubits, they need to be prepared in a (pseudo-)pure state. The remaining subsystems  $\mathcal{S}_i$ can however remain in a mixed state. We require only that the logical state evolves as a pure state in the logical subsystem, under the action of logical operations.  

If we are evolving the system with logical operators, the fact that they act only on the encoded subspace $\mathcal{L}$  ensures that information within this subspace will not leak out  or mix with the orthogonal spaces during logical unitary transformations, thus preserving the purity of the encoded subspace under the noise model.
An important requirement for the subsystem pseudo-pure states is the ability to decode: the use of a mixed state should not introduce a mixing of the  information contained in the logical qubits and in the unprotected subsystems, even when the information is transfered back to physical qubits by decoding. For unital maps, setting the unprotected subsystem to the identity state will satisfy this requirement without any further action required on the decoded state (notice that other mixed states 
are possible for particular encodings).		
For a DFS or a NS, not being able to apply the simple decoding operation to transfer  the full information back to the physical qubits is inconvenient, as logical observables are in general difficult to measure experimentally since they are usually given by many-body states in the basis of the physical system. In the case of QEC the decoding involves also a correction step. If the unprotected part of the Hilbert space has evolved, it is no longer possible to perform a unique correction operation, valid for any input state.

The state preparation procedure that bears the most resemblance to the method we propose is logical labeling \cite{ChuangPP}, which uses a unitary transformation to change the equilibrium distribution of spin states into one where a subsystem of the Hilbert space is pseudo-pure conditional on a physical spin having some preferred orientation.  The parts of Hilbert space that remain mixed are of no use to the computation.  It can be shown that a $m-$qubit  pseudo-pure state can be stored among the Hilbert space of $N-$qubits provided the inequality $(2^m-1) \leq \frac{N!}{((N/2)!)^2}$ is satisfied.  A key insight is that in the study of encoded qubits, one need not take this $m-$qubit effective pure state and perform an encoding of $l-$logical qubits under the hierarchy $l < m < N$: Instead, a $l$-qubit encoded state can be prepared directly from the equilibrium state of $N$ qubits.

If information is encoded in a subsystem of dimension $2^l$, with a corresponding syndrome subsystem, $\mathcal{S}$ of dimension $d_s$ that we can leave in a mixed state, the number of zero eigenvalue in this subsystem pseudo-pure state is $(2^l-1)d_s$. We can create a state that is pure on the logical degrees of freedom as long as there are at least as many zero eigenvalues in the thermal state as in the $l$-qubits pure state: $(2^l-1) \leq \frac{N!}{((N/2)!)^2 d_s}$.  
However, the eigenvalue spectrum of the equilibrium state of an $N$ spin density matrix ( $\lambda(\rho_{eq}) = \{N,N-2,...,-N\}$) will most generally not generate the necessary eigenvalue spectrum required for decoding the $l$-qubits of information into $l$ physical qubits without error. So a combination of techniques must be used.

Before presenting a general model that allows us to quantify the SNR gain obtained by the subsystem pseudo-pure states, we will clarify the concept with an example.

\section{Example}
In a DFS information is protected inside subspaces of the total Hilbert space that are invariant under the action of the noise generators.
Here we consider  collective $\sigma_z$ noise, which describes a dephasing caused by completely correlated fluctuations of the local magnetic field $B_z\hat{z}$: 
\begin{equation}
	\label{Noise}
	\mathcal{H}_{SE} = J_z \otimes B_z, 
\end{equation}
with $J_z = \sum_i \sigma_z^i$, the total spin angular momentum along the quantization axis $z$.
In a Hilbert space of dimension 4, the eigenspace of the noise operator $J_z$ with eigenvalue 0 is a 2-dimensional DFS \cite{DFSEvan} and can be used to encode one qubit of information. 
The DFS is spanned by the basis vectors $|01\rangle$ and $|10\rangle$.  A natural encoding of a logical qubit $|\psi\rangle_L$ is given by:
\begin{equation}
	\label{encoding1}
	 \alpha |0\rangle_L + \beta |1\rangle_L \Longleftrightarrow  \alpha |01\rangle + \beta |10\rangle
\end{equation}

	 The encoded pure state for a DFS logical qubit is given by:
\begin{equation}
\label{DFSPP}
\begin{array}{l}
\displaystyle	 \ket{01}\bra{01}
					=\frac{\openone_L+\sigma_{z,L}}{2} = \frac{\openone + \sigma_z^1 - \sigma_z^2 - \sigma_z^1\sigma_z^2}{4} 
					\end{array}
\end{equation}
In the case of this DFS, the Hilbert space can be written as a direct sum of the logical subspace $\mathcal{L}$ (spanned by the basis $|01\rangle$ and $|10\rangle$) and its complementary subspace $\mathcal{R}$ (spanned by the basis $|00\rangle$ and $|11\rangle$). If we add the identity on the $\mathcal{R}$ subspace to the logical pure state, we obtain a mixed state that is equivalent in terms of its behavior on the logical degrees of freedom: 
\begin{equation}
	\rho = \half|0\rangle\langle 0|_L+\frac{\openone_R}{4}  =\frac{1}{4}(\openone+\sigma_{z,L}) =\frac{1}{4}(\openone + \frac{\sigma_{z}^1-\sigma_z^2}{2})
\end{equation}
 The traceless part of this state is simply $\propto \sigma_z^1-\sigma_z^2$: From thermal equilibrium, a unitary operation is sufficient to obtain this state, so no signal is lost. 
The subsystem pseudo-pure state that one obtains with this method requires less  averaging to implement the non-unitary transformation. We expect such transformation to result in general  in  higher SNR, since less information about the system is neglected, and to have less complex state preparation, since the non-unitary transformation is less demanding.

As an example, consider the pure state of two logical qubits encoded into a 4 physical qubit DFS \cite{viola2encdyndec}:
\begin{equation}
\label{logicalPP}
\begin{array}{l}
	\ket{00}\bra{00}_L =\frac{1}{4}(\openone_L^{1} + \sigma_{z,L}^{1})\otimes (\openone_L^{2} + \sigma_{z,L}^2)
					\end{array}
\end{equation}
If we add $\openone_{R}$ to the unprotected subspace of each logical qubit we obtain a state which is pseudo-pure within the subspace of the logical encoding:

\begin{equation}
\label{logicalPP2}
\begin{array}{lll}
	\rho_{ep} & = &\frac{1}{16}(\openone_L^1 +\openone_R^1+ \sigma_{z,L}^1)\otimes (\openone_L^2+\openone_R^2 + \sigma_{z,L}^2)\\ \\
					& = &\frac{12}{16}(\openone^{1,2} + \frac{\sigma_z^1 - \sigma_z^2}{2})\otimes(\openone^{3,4} + \frac{\sigma_z^3 - \sigma_z^4}{2}) 
					\end{array}
\end{equation}

We still need a non-unitary operation to obtain this state, but the resulting SNR  is  $2/3$ instead of $4/15$ as for creating the full pseudo-pure state. 
The preparation procedure is also less complex, since it only requires preparing up to $2-$body terms ($\sigma_z^i\sigma_z^j$) instead of the $4-$body term ($\sigma_z^1\sigma_z^2\sigma_z^3\sigma_z^4$) necessary for the full pseudo-pure state (in general  an N-body term involves interactions among all N spins; usually only some of the couplings among spins are strong enough to permit fast two-qubit operations). 

\section{General Theory}
 To illustrate the advantages that subsystem pseudo-pure states bring, we now present  a general scheme, looking for a quantitative bound on the increase in sensitivity with respect to  full pseudo-pure states.
When under a particular encoding the Hilbert space is transformed to $\ham \cong \bigoplus_i \mathcal{L}_i \otimes \mathcal{S}_i$, in the encoded representation the state we want to prepare will have the form:
\begin{equation}
	\rho_{spps}= \bigoplus_i a_i \left(|\psi\rangle_L^i\langle\psi|^i_L \otimes \frac{\openone_S^i}{d_{s_{i}}}\right)
\end{equation}
The dimension of the $i$-th syndrome is $d_{s_{i}}$; $a_i$ are subspace weighting coefficients, such that $\sum_i a_i = 1$, ensuring a unit trace of $\rho_{spps}$.
We  analyze the conditions that  give the optimal signal, given that some freedom in the construction of the subsystem pseudo-pure states is available. 

Since we are interested in the information that we can manipulate and observe, a good measure of the sensitivity gain is the SNR of the qubits storing the information after the decoding. Instead of the total magnetization, which is the observable in NMR, we are therefore interested in: \begin{equation}\label{SNR}\begin{array}{l}
	SNR =\langle |\vec{M}|\rangle \propto S(\rho)=
	\sqrt{\Tr{\sum\sigma_z^i \rho}^2+\Tr{\sum\sigma_x^i \rho}^2+\Tr{\sum\sigma_y^i \rho}^2}
\end{array}\end{equation}
where the sum only extends over the $l$ information carrying qubits. 
Other metrics are of course conceivable, for example the total magnetization of the $N$ spins or the spectral norm of the density matrix deviation, but they are not directly related to the signal arising from the information carrying qubits \footnote{Notice that even the chosen metric can be misleading, since in the case $l>1$ some states give no signal (e.g. a non-observable coherence). However, any of these states can be characterized by especially designed read-out operations that transform it to an observable state, while preserving its information content. More specifically, we will consider only the ground state $\ket{00\dots}$ signal, since any other state is isomorphic to it, via a unitary operation.}.

We assume to encode $l$ logical qubits among $N$ physical qubits, with a syndrome subsystem $\mathcal{S}$ of dimensions $2^s$ \footnote{We consider only the case where we can map qubits on the subsystem S, even if in general the subsystem could not be mapped to qubits. The results would be however the same, with slightly different notations.} (the Hilbert space can be written as $\ham=\mathcal{L}\otimes\mathcal{S}\oplus\mathcal{R}$) . 
The encoding operation is, in general, defined by its action on the initial state $\ket{\psi}_l\ket{00\dots}_{N-l}$, giving the encoded state: $\ket{\psi}_L\ket{0}_S$. Hence, there is some flexibility in the choice of encoding operation (since it is defined only for ancillas initially in the ground state) but we can specify it with the assumption that the state in the encoded subsystem $\mathcal{S}$ is determined by the first $s$ ancillas state:
\begin{equation}
	\label{DecodingDefinition}
	U_{enc}\ket{\psi}_l\ket{\phi}_s\ket{00...}_{N-l-s}=\ket{\psi}_L\ket{\phi}_S
\end{equation}

The subsystem pseudo-pure state is:
\begin{equation}	\rho_{spps}=a \left( \ket{\psi}\bra{\psi}_L\otimes\frac{\openone_S}{2^s} \right)+\frac{1-a}{2^N-2^{s+l}}\openone_R,
\end{equation}
which after decoding following (\ref{DecodingDefinition}) becomes:
\begin{equation}
\label{NSdec}\begin{array}{l}	\rho_{spps}=\left(a\ket{\psi}\bra{\psi}-\frac{1-a}{2^N-2^{s+l}}\openone_l\right)\frac{\openone_s}{2^s}\ket{00\dots}\bra{00\dots}_{N-l-s}\\
+\frac{1-a}{2^N-2^{s+l}}\openone,
\end{array}\end{equation}
so that the signal is given by: $S(\rho)=a\epsilon\alpha S(\ket{\psi}\bra{\psi}_l)\propto a\alpha l$, where $\alpha\leq \frac{\left\| \rho_{eq} \right\|}{\left\|\rho_{pp}-\openone/2^N\right\|}$. 
To obtain the spectral norm of the subsystem pseudo-pure state traceless part, we calculate its eigenvalues:
\begin{equation}\label{eigNS}
	\{\frac{a}{2^s}-2^{-N},\ -2^{-N},\ \frac{1-a}{2^N-2^{s+l}}-2^{-N}\}
\end{equation}

The upper bound for the signal is  obtained for $a=2^{s+1-N}$ and we have: $SNR \propto N2^{s+1-N}$.

For more than one logical qubit, each being protected against some noise, or  to concatenate different encodings, a tensor structure of encoded qubits arises naturally. We analyze also this second type of construction, that can bring a further enhancement of the signal. We assume here that we encode one logical qubit in $n$ physical qubits -each being a subsystem pseudo-pure state- and we build a logical $l-$qubit state with the tensor product of these encoded qubits.
The Hilbert space can be written as a tensor product of direct sums as: $\ham=\bigotimes_{i=1}^l(\mathcal{L}_i\otimes\mathcal{S}_i\oplus\mathcal{R}_i)$. 

The corresponding partially mixed states differ with respect to the previous ones, in that subspaces that are not actually used to store protected information are not maximally mixed:
\begin{equation}\label{DFSprod}\begin{array}{l}	\rho_{spps}=(a_1\ket{\psi}\bra{\psi}_{L1}\otimes\frac{\openone_{S1}}{2^s}\oplus\frac{1-a_1}{2^n-2^{s+1}}\openone_{R1})\\ \otimes(a_2\ket{\psi}\bra{\psi}_{L2}\otimes\frac{\openone_{S2}}{2^s}\oplus\frac{1-a_1}{2^n-2^{s+1}}\openone_{R2})\otimes\dots
\end{array} 
\end{equation}
 Notice that we consider  that all the qubits have the same encoding and to make the problem more tractable we  choose $a_i=a, \forall i$. Other choices of course exist, and may lead to a better SNR, but will not be explored here.
Upon decoding, this state is transformed to:
 \begin{equation}\label{DFSdecprod}\begin{array}{ll}
	U_{enc}^\dag\rho_{spps}U_{enc} & =\bigotimes_{i=1}^l[(a\ket{\psi}\bra{\psi}_1-\frac{1-a}{2^n-2}\openone_1)\\&  \otimes \frac{\openone_s}{2^s}\otimes  
  \ket{00\dots}\bra{00\dots}+\frac{1-a}{2^n-2}\openone_n]_i
\end{array}\end{equation}

The signal is again proportional to $a\alpha l$ and varying $a$ we can find the optimal state. 
The eigenvalues for the traceless part of the subsystem pseudo-pure state are:
\begin{equation}\label{eigProd}\begin{array}{c}
  \prod_{i=1}^l (\{\frac{a}{2^s},0,\frac{1-a}{2^n-2^{s+1}}\}_i)   -2^{-N} \\  = \{(\frac{a}{2^s})^{l-p}\left(\frac{1-a}{2^n-2^{s+1}}\right)^p\|_{p=0}^l-2^{-N}, -2^{-N}\}
\end{array}\end{equation}

 The maximum SNR depends on the relative dimension of the logical subspace and the syndrome and on the number of encoded qubits. In particular:

When $l < \frac{-1}{\log_2(1-2^{s-n})}$ the $SNR\propto N2^{s+1/l-n}$ (The norm reaches the minimum value $2^{-N}$ for $a\leq 2^{s+1/l-n}$). 
 
 When $l > \frac{-1}{\log_2(1-2^{s-n})}$, instead, we obtain $SNR \propto \frac{N2^s(2^n-2^s)^l}{2^N-(2^n-2^s)^l}$: The minimum value for the norm $(2^n-2^s)^{-1}-2^{_N}$ is obtained for $\frac{a}{2^s}=\frac{1-a}{2^n-2^{s+1}}$, i.e. for $a=\frac{2^s}{2^n-2^s}$). 
 
  Notice that for $n\leq s$ there is no useful solution. 
 In both cases, the SNR obtained with a tensor product structure is higher than for the first construction presented.

\begin{figure}[hbt]
	\centering
		\includegraphics[scale=0.3]{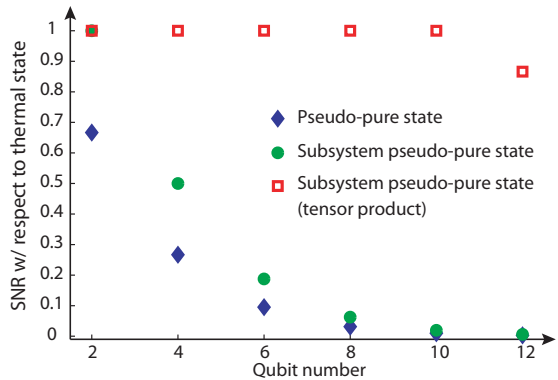}
	\label{fig:DFSSNR}
	\caption{SNR (normalized to the SNR for thermal state) for DFS encoding of 1 logical qubit on two physical qubits, as a function of the total physical qubit number.} 
\end{figure}

The improvement brought by the subsystem pseudo-pure states can be generalized to many types of encoding. We now present two examples, applying our scheme to the case of a 3-qubit NS and a 3-qubit QEC, to illustrate some possible applications of the scheme proposed.

\textbf{Noiseless Subsystems.}  
The smallest code that protects a system against an arbitrary collective noise is realized with a 3 physical qubit NS \cite{noiseless,DFSkempe,noiselessSC,QECNS,BlattNS}. The collective noise conserves the total angular momentum $J$ of the system. In the case of 3 spin-$\half$ system, the Hilbert space can immediately be written as: $\mathcal{H}=\mathcal{H}_{3/2}\oplus\mathcal{H}_{1/2}$. The second subspace ($\mathcal{H}_{1/2}$)is doubly degenerate, so we  identify in it a protected subsystem, reflecting a logical degree of freedom: $\mathcal{H}_{1/2}=\mathcal{L}\otimes \mathcal{S}$, where the second subsystem is associated with the $j_z$ quantum number. 

 Since the information is all encoded in the subsystem $\mathcal{L}$, we can safely leave the subsystem $\mathcal{S}$ in a mixed state. The state we want to create has  the form: $|0\rangle\langle0|_L\otimes \openone_S/2$. In terms of physical operators, using the decoding operator in ref. \cite{noiseless}, this corresponds to: 
\begin{equation}
	\label{NSLP}
	\rho_{spps}=(\openone+\sigma_z^1+\sigma_z^2+\sigma_z^1\sigma^2_z)/8
\end{equation}
Only $1/3$ of the equilibrium signal must be lost to create this state.

Notice that we can also set the subspace corresponding to $j=3/2$ to the identity state, with $a=\half$ for the optimal SNR. With the encoding given in \cite{noiseless}, the identity on the unprotected subspace $\mathcal{H}_{3/2}$ is $(\openone-\sigma_z^1/2)/8$. 
The subsystem pseudo-pure state is then: $\rho_{spps}=(2\cdot\openone+\sigma_z^1/2+\sigma_z^2+\sigma_z^1\sigma^2_z)/16$ and $1/2$ of the SNR is retained in obtaining this state instead of $3/7$ for a full pseudo-pure state, a $16\%$ increase.

\textbf{QEC.}
When the noise does not present any useful symmetry, information can still be preserved by using QEC codes \cite{shor95,steane96,CalderbankShor96}. These codes are based on a two step operation for protecting against the noise: first the information is encoded in an appropriate subspace and then, after a short time during which at most one error has occurred, the qubit is corrected based upon the state of the syndrome. 
To obtain a subsystem pseudo-pure state, the strategy is again to purify only the subspace encoding the information, while the ancillary subspaces are left in a mixed state. It could appear that this scheme  cannot be applied, since when ancillas are not in the ground state but in a mixed state, they indicate that errors had already acted on the system: QEC codes can protect only for a finite number of errors. However, if we initially populate the orthogonal syndrome subspaces with the identity, recovery of the information is still possible and we obtain an subsystem pseudo-pure state with a higher SNR and the same observable dynamics as a full pseudo-pure state.

Consider for example the encoding for the 3 qubit QEC \cite{SteaneQEC,coryQEC}, protecting against a single bit-flip error ($\sigma_x^i$). The code subspace is spanned by the basis set:
\begin{equation}
	\label{QEC}
	\begin{array}{ll}
	|0\rangle_L=|000\rangle ;&\ \ |1\rangle_L=|111\rangle  
	\end{array} 
\end{equation}

The Hilbert space has an irreducible representation as the direct sum of 4 orthogonal subspaces:  $\mathcal{H}=\mathcal{L}\oplus\mathcal{R}_1\oplus\mathcal{R}_2\oplus\mathcal{R}_3$, each $\mathcal{R}_i$ spanned by the basis: $\sigma_x^i\{\ket{0}_L,\ket{1}_L\}$, $i=1,2,3$. 

An error causes a swapping of the code subspace with one of the orthogonal subspaces, which is then corrected by the decoding operation. Starting from the pure state: $\ket{\psi}\ket{00}$, which we encode following (\ref{QEC}), the final state after an error and decoding is $\ket{\psi}\ket{xy}$ ($x,y \in\{0,1\}$) and the ancillas need to be reinitilized for the code to be effective against a second error.
Since logical operations act only on the first subspace, we can set the other subspaces to the maximally mixed state. The state we want to prepare is thus given by:
\begin{equation}	\label{QECLP}
\begin{array}{l}
	a\ket{\psi}\bra{\psi}_L+(1-a)(\openone_{R1}+ \openone_{R2}+ \openone_{R3})/6 \\ = a\ket{\psi}\bra{\psi}_L+(1-a)(\openone- \openone_L)/6  
	\end{array}
\end{equation}
When we decode after an eventual error, we obtain the state: $(a\ket{\psi}\bra{\psi}_1-(1-a)\openone_1/6)\ket{xy}\bra{xy}+(1-a)\openone/6$.
Since the identity $\openone_1$ is not observable in NMR, this state carries the same information content as the full pseudo-pure state.
With $a=1/4$ we find that the SNR is reduced only to $\frac{3}{4}SNR$ for this mixed state, while it would be $3/7$ for the full pseudo-pure state. 

Generalizing to other QEC codes, one can always find an encoding operation that transform the Hilbert space to $\ham \cong \mathcal{L} \bigoplus_i \mathcal{R}_i = \mathcal{L} \oplus \mathcal{R}$ and prepare a subsystem pseudo-pure state following the constructions for the DFS, setting the subspace $\mathcal{R}$ to identity. 
However, this encoding only allows one to correct for a finite number of errors, a recovery operation is needed to reinitialize the ancillas.  The recovering operation could be in general accomplished by a strong measurement, however this is not feasible in NMR; one must have fresh ancillas available (the problem posed by the need of resettable ancilla is not unique to NMR). 
To correct for two errors in the previous example, a partially mixed state with 4 ancillas should be prepared, so that two new fresh ancillas can be used for correcting the second error. In general, in addition to the 3 qubit system that encode the state, a separated reservoir (not affected by the noise) of $2n$ ancillas is needed for correcting $n$ errors. 
 Even if ancillas must be prepared all simultaneously, the creation of subsystem pseudo-pure states increase the SNR, so that the number of ancillas, and therefore of error that can be corrected, can be  increased in actual experiments.

\section{Metrics of control}
The correlation of the experimental density matrix with the theory density matrix is a quantitative measure of control \cite{softpulses}. The attenuated correlation takes into account attenuation due to decoherent or incoherent processes:

\begin{equation}
	C = \frac{\Tr{\rho_{th} \rho_{exp}}}{\sqrt{\Tr{\rho_{th}^2}\Tr{\rho_{in}^2}}}.
\end{equation}
Here $\rho_{th}=U\rho_{in}U^\dag$, $\rho_{exp} = \mathcal{E}(\rho_{in})=$, and $\rho_{in}$ define the theoretical, experimental and input states respectively. 

When we compare theoretical and experimental encoded states, their overlap  has contributions that mirror the logical subsystem structure of the Hilbert space. Consider for simplicity a Hilbert space that can be written in terms of a logical and non-logical subspaces, $\ham\cong\mathcal{L}\oplus\mathcal{R}$. Rewriting the experimental quantum process in terms of Kraus operators \cite{Kraus} $A_\mu$, ($\mathcal{E}(\rho) = \sum_\mu A_\mu \rho A_\mu^\dag $) we can separate them into three groups: $\{A_{\mu,\mathsf{LL}},A_{\mu,\mathsf{RR}},A_{\mu,\mathsf{LR}}\}$, which respectively describe the maps on the $\mathcal{L}$ subspace, $\mathcal{R}$ subspace, and the mixing of these two subspaces. The correlation will reflect these three contributions to the dynamics, $C = \alpha_{\mathsf{LL}}C_{\mathsf{LL}} +(\alpha_{\mathsf{LR}}C_{\mathsf{LR}}+\alpha_{\mathsf{RL}}C_{\mathsf{RL}})+\alpha_{\mathsf{RR}}C_{\mathsf{RR}}$, where:  
\begin{equation}\begin{array}{ll}
   \displaystyle
	 C_{\mathsf{KH}}=\frac{\Tr{P_\mathsf{K}\rho_{th} \sum_\mu A_{\mu,\mathsf{KH}} (P_\mathsf{H}\rho_{in}P_\mathsf{H})A_{\mu,\mathsf{KH}}^\dag}}{\sqrt{\Tr{(P_\mathsf{H}\rho_{in})^2}\Tr{(P_\mathsf{K}\rho_{th})^2} }},\ \ \ \ \ 
	  & \alpha_{\mathsf{KH}}=\sqrt{
	  	\frac{
			\Tr{
				(P_\mathsf{H}\rho_{in})^2 (P_\mathsf{K}\rho_{th})^2
			     }
		        }
		        {
		        	\Tr{\rho_{in}^2}\Tr{\rho_{th}^2}
			}
		}
\end{array}
\end{equation}

Here we define $P_{\mathsf{L}}$ ($P_{\mathsf{R}}$)  as the projector onto the encoded (non-logical) subspace. Notice that if the ideal state is inside the logical subspace, its projection on the non-logical subspace $P_\mathsf{R}\rho_{th}$ is zero and the last term goes to zero, $C_\mathsf{RR}=0$. 

For encoded qubits, we limit state tomography to the logical subspace only, so that a reduced number of readouts is enough to characterize the information available from the this subspace. The ability to preserve and manipulate the information inside the logical subspace can be better quantified by the correlation on this subspace, $C_\mathsf{LL}$, comparing the experimental logical state with the theoretical state inside the subspace only.  
If the input state of this process, $\rho_{in}$, is a full pseudo-pure state and inside the logical subspace, the correlation $C_{\mathsf{LL}}$ with the logical ideal state is the only contribution to the total correlation $C$.  

If a pseudo-pure state over logical degrees of freedom is used instead, $C_{\mathsf{LR}}\neq 0$, since  the output state in the protected subspace may contain contributions arising from the action of the map $\mathcal{E}$ on the identity in the non-logical subspace. 
Given an input state $\rho_{in}=a\ket{\psi}\bra{\psi}_L+(1-a) \frac{\openone_R}{d_R}$, from the experimental output state we can only measure the quantity (by observing only the $l$-logical qubits or their phyisical equivalents):
\begin{equation}
C^*_{\mathsf{LL}} = a C_{\mathsf{LL}}+(1-a)C_{\mathsf{LR}}=a C_{\mathsf{LL}}+(1-a)\frac{\Tr{U_{th}\ket{\psi}\bra{\psi}_LU^\dag_{th} \mathcal{E}\big(\frac{\openone_R}{d_R}\big)} } {\sqrt{\Tr{(P_L\rho_{in}P_L)^2}}\sqrt{\Tr{(P_L\rho_{th}P_L)^2}}}
\end{equation}
 Note that in this case $C \neq C^*_{\mathsf{LL}}$, since the contribution $C_{\mathsf{RL}}$ is not taken into account.
The measured correlation is thus defined by two terms: the first takes into account the control over the encoded subspace only and the eventual leakage from it, while the second takes into account mixing from the $\mathcal{R}$ subspace to the $\mathcal{L}$ subspace. State tomography of the  input state $\openone_R$ after the algorithm allows one to calculate the correlation on the logical subspace $C_{\textsf{LL}}$.

To characterize the control of quantum gate operations most generally, many metrics have been suggested \cite{softpulses, SchumFid,  NielsenFid}. A good operational metric is for example the average gate fidelity (or fidelity of entanglement), that can be measured as the average of correlations of a complete orthonormal  set of  input states: $\bar{F}=\sum_j C^j=\sum_j \Tr{U_{th}\rho^jU_{th} \mathcal{E}(\rho^j)}$. 
Similarly, the encoded operational fidelity can be defined as the average correlation over an orthonormal set of operators spanning $\mathcal{L}$: $\bar{F}_L=\sum'_j C_{LL}^j=\sum'_j\Tr{U_{th}\rho_L^jU_{th} \mathcal{E}(\rho_L^j)}$. 

The fidelity on the logical subspace focuses on the achieved control in the implementation of the desired transformation on the protected subspace; this new metric is immune to unitary or decoherent errors within $\mathcal{R}$ alone: 
\begin{equation}
\begin{array}{l}
\bar{F}_L=\sum'_j C_{LL}^j=
\sum'_j\Tr{U_{th}\rho_L^jU_{th} (\sum_\mu A_\mu P_L\rho_L^jP_LA_\mu) } \\=\sum_\mu |U_{th}A_{\mu,L}|^2/N^2
\end{array}\end{equation}

The extent to which $U_{L, exp}$ is close to $U_{th,L}$ can be determined from $\bar{F}_L$, while the avoidance of subspace mixing will be specified by the gap between $\bar{F}$ to $\bar{F}_L$. %

\section{Conclusions}
	Subsystem pseudo-pure states offer not only a greater SNR but also a less complex state preparation which is reflected in increased fidelity of the  experimental results (as shown in a following paper).  By no means does this logical encoding overcome the exponential loss of signal suffered by pseudo-pure states; however, for the corresponding state in the full Hilbert space, the gain is significant.  As we explore control over multiple logical qubits, these advantages become tantamount.
\begin{table}[htb]
\small
	\centering
	\begin{tabular}{c|m{1.2cm}|l}
			\hline
			& & \vspace{-.3cm}\\
			\multirow{2}* {\textbf{4  spins}} 
					& \centering \multirow{2}*{DFS}  &1 logical qubit. Isotropic noise. \\ 
					& & What differences with 3-qubit NS? 
 			\\[.1cm]\hline\hline
 			& & \vspace{-.3cm}\\
			\multirow{3}* {\textbf{5 spins}} & 
			\centering {NS} &	 2 logical qubits.  Create Bell-State \\[.1cm] \cline{2-3}
			& & \vspace{-.3cm}\\
			&
			\centering {QEC}	&  2 errors. 1-logical qubit QEC for $\sigma_x$ errors.  \\[.1cm] \cline{2-3}
			& & \vspace{-.3cm}\\
			&
			\centering {QEC}	&  2-logical qubits QEC for $\sigma_x$ errors
			\\[.1cm]\hline\hline
			& & \vspace{-.3cm}\\
			\multirow{4}* {\textbf{6 spins}} & 
				\centering {DFS} & 3-logical qubits  $\sigma_z$ noise.  Create GHZ state \\[.1cm]  \cline{2-3}
				& & \vspace{-.3cm}\\
			& 	\centering {GHZ} &    Create 2 GHZ then encode them.   $\sigma_z$-noise \\[.1cm]  \cline{2-3}
			& & \vspace{-.3cm}\\
			& \centering {QEC DFS} & Concatenate 2-qubit DFS with 3-qubit QEC.
			\\[.1cm]\hline\hline
			& & \vspace{-.3cm}\\
			{\textbf{9  spins}} & \centering {Shor's Code} & 9-qubit QEC to correct all single-qubit errors \\[.1cm]\hline			
		\end{tabular}
	\caption{In the table are shown experiments on 4-9 qubits that will be achievable with the initialization method proposed}
	\label{EncodedExp}
\end{table}
	
	In particular, this method coupled with the experimental control on an Hilbert space of about 10 qubits, would allow for the study of a repeated QEC code, the Shor code for protecting against single qubit errors, or a multi-layered encoding like QEC using three DFS encoded qubits or vice-versa.  In addition encoded versions of gates essential to algorithms, like the Quantum Fourier Transform, can be carried out with liquid phase NMR. The effective noise superoperator on the logical subspace can also be reconstructed \cite{WeinsteinQFT}, to gain an insight on how various encoding schemes modify the noise structure.
	
	By considering metrics of control for only the logical degrees of freedom of our system, we also reduce the number of input states needed to characterize a particular gate sequence, as far as only the behavior of the protected information is of interest.  

The state initialization method proposed could also find application in a broader context, whenever exact purification of the system is possible, but costly. Quite generally, a qubit need not to be identified with a physical two-level system, but rather with a subsystem whose operator algebra generators satisfy the usual commutation and anti-commutation relationships. State initialization and purification could be performed  on these subsystems only, thus allowing experimental advantages similar to the ones shown in the particular case of logical qubits.

\textbf{Acknwoledgements}: The authors thank N. Boulant and P. Zanardi for helpful discussions.
This work was supported in part by the National Security Agency (NSA) under Army Research Office (ARO)
contract numbers  W911NF-05-1-0469 and DAAD19-01-1-0519, by the Air Force Office of Scientific Research, and by the Quantum Technologies Group of the Cambridge-MIT Institute.
\bibliography{SubsystemPPS}
\end{document}